\documentclass[twocolumn,showpacs,preprintnumbers,amsmath,amssymb,aps,prl]{revtex4-1}

\usepackage{graphicx}
\usepackage{dcolumn} 
\usepackage{bm, bbm}      
\usepackage{curves}
\usepackage{epic}
\usepackage{wasysym}
\usepackage{sidecap}
\usepackage{multirow}
\usepackage{amssymb, amsfonts, latexsym}
\usepackage[mathcal]{euscript}
\usepackage{epsf, times}
\usepackage{subfigure}
\usepackage{amsthm}

\def\openone{\leavevmode\hbox{\small1\kern-4.2pt\normalsize1}}

\def\S{\mathcal{S}}
\def\A{\mathcal{A}}
\def\B{\mathcal{B}}
\def\Ai{\mathcal{A}_1}
\def\Aii{\mathcal{A}_2}
\def\Bi{\mathcal{B}_1}
\def\Bii{\mathcal{B}_2}

\def\N{\mathcal{N}}
\def\E{\mathcal{E}}

\def\Id{\mathbb{I}}
\def\og{\overline{g}}
\def\oog{\overline{\overline{g}}}
\def\tg{\tilde{g}}

\newcommand{\beq}{\begin{equation}}
\newcommand{\eeq}{\end{equation}}
\newcommand{\bea}{\begin{eqnarray}}
\newcommand{\eea}{\end{eqnarray}}
\newcommand{\bfig}{\begin{figure}}
\newcommand{\efig}{\end{figure}}

\begin{document}

\title{
Negativity and topological order in the toric code
      }

\author{
C. Castelnovo$^{1}$
}
\affiliation{
$^1$
TCM group, 
Cavendish Laboratory, 
University of Cambridge, 
Cambridge CB3 0HE, United Kingdom
}

\date{\today}

\begin{abstract}

In this manuscript we study the behaviour of the entanglement measure 
dubbed negativity in the context of the toric code model. 
Using a replica method introduced recently by Calabrese, Cardy and Tonni 
[Phys. Rev. Lett. 109, 130502 (2012)], we obtain an exact 
expression which illustrates how the non-local correlations present in a 
topologically ordered state reflect in the behaviour of the negativity 
of the system. 
We find that the negativity has a leading area-law contribution, if the 
subsystems are in direct contact with one another (as expected in a 
zero-range correlated model). We also find a topological contribution 
directly related to the topological entropy, provided that the partitions 
are topologically non-trivial in both directions on a torus. 
We further confirm by explicit calculation that 
the negativity captures only quantum contributions 
to the entanglement. Indeed, we show that the negativity vanishes identically 
for the classical topologically ordered 8-vertex model, which on the contrary 
exhibits a finite von Neumann entropy, inclusive of topological correction. 

\end{abstract}

\maketitle
%
%

\section{Introduction} 
In the effort to understand and quantify classical vs quantum correlations in 
many body systems, a number of different measures of entanglement have been 
proposed in recent years. The von Neumann entropy, for instance, is obtained 
from a bipartition of the system $\S = \A \cup \B$: 
$S_{\rm vN}^{(\A)} = - {\rm Tr} \rho_\A \ln \rho_\A$, where 
$\rho_\A = {\rm Tr}_\B \rho$. 
This definition, however, is a measure of quantum correlations between $\A$ 
and $\B$ only if $\rho$ is a pure state. 
In order to apply it to mixed states, one ought to symmetrise it and compute 
the mutual information 
$S_{\rm vN}^{(\A)} + S_{\rm vN}^{(\B)} - S_{\rm vN}^{(\A\cup\B)}$, 
which nonetheless remains sensitive to classical as well as quantum 
correlations, and it is therefore only an upper bound on the entanglement 
between $\A$ and $\B$. 

Providing an explicit measure of entanglement that applies to both mixed and 
pure states, and that is of practical use, has been a tall order. 
In recent years it was proposed to use of a quantity called 
\emph{negativity}, which was first introduced in 
Ref.~\onlinecite{Zyczkowski1998}, and later proven to be an entanglement 
monotone by several authors~\cite{Lee2000,Eisert2001,Plenio2005,Vidal2002}. 
The negativity $\N$ (or, equivalently, the \emph{logarithmic negativity} 
$\E$), is defined from the trace norm $\| \rho^{T_\B} \|_1$ of the 
partial transpose over subsystem $\B$ of the density matrix $\rho$, 
\bea
\N &\equiv& \frac{\| \rho^{T_\B} \|_1 - 1}{2} 
\\
\E &\equiv& \ln \| \rho^{T_\B} \|_1
, 
\label{eq: log negativity}
\eea
where $\| \rho^{T_\B} \|_1$ is the sum of the absolute 
values of the eigenvalues $\lambda_i$ of $\rho^{T_\B}$. 
If all the eigenvalues are positive then $\N = 0$ (recall that 
$\sum_i \lambda_i = 1$), and $\N > 0$ otherwise -- hence its name. 
The existence of negative eigenvalues is directly related to the fact that 
$\A$ and $\B$ are not separable, as discussed e.g., in 
Ref.~\onlinecite{Vidal2002}. 

Despite the availability of an explicit formulation, 
the calculation of the negativity of a 
many body quantum system remains an arduous task which has been carried 
out in the literature mostly in 1D. 
Recently, Calabrese, Cardy and Tonni~\cite{Calabrese2012} devised a new 
scheme based on a replica approach, which allows to compute the negativity 
in conformally invariant field theories 
(see also Refs.~\onlinecite{Calabrese2013,Alba2013}). 

Here we apply the method introduced in Ref.~\onlinecite{Calabrese2012} to 
compute the negativity of the toric code model~\cite{Kitaev2003}. 
We are able to obtain an exact expression which illustrates the microscopic 
origin of the different contributions to the negativity, depending on 
the nature of the partition of the system. 
In addition to the expected area-law contribution if the subsystems are in 
direct contact with one another, we also find that the non-local correlations 
present in a topologically ordered state affect the behaviour of the 
negativity. 
This {\it topological} contribution relates directly to the topological 
entropy and appears only if the partitions are topologically 
non-trivial, which is consistent with the fact that topologically trivial 
disconnected subsystems are separable in a topologically ordered state. 

With this calculation we also show that the negativity captures only the 
off-diagonal (`quantum') contribution to the topological 
entropy~\cite{Levin2006,Kitaev2006} and it 
is insensitive to the diagonal part. Indeed, we find 
that $\N=0$ for the classical topologically ordered 8-vertex model 
(which on the contrary has a non-vanishing topological 
entropy~\cite{Castelnovo2007}). 
%
%

\section{The toric code model} 
The toric code is a system of spin-1/2 degrees of freedom 
$\sigma_i$ living on the bonds $i$ of a square lattice (periodic boundary 
conditions will be assumed throughout). 
The Hamiltonian of the system can be written as~\cite{Kitaev2003}: 
\bea
H = - \lambda_A \sum_s A_s - \lambda_B \sum_p B_p 
, 
\label{eq: TC Hamiltonian}
\eea
where $s$ ($p$) label the sites (plaquettes) of the lattice, and 
$A_s = \prod_{i \in s} \sigma^x_i$, $B_p = \prod_{i \in p} \sigma^z_i$. 

The ground state (GS) is 4-fold degenerate, according to the 4 topological 
sectors identified by the expectation value of winding loop operators. 
Within each sector, the GS is given by the equal amplitude superposition of 
all tensor product basis states $\otimes_i \vert \sigma_i^z \rangle$ 
belonging to that sector. 
Following the notation in Refs.~\onlinecite{Hamma2005,Castelnovo2008}, 
we define $\vert 0 \rangle \equiv \otimes_i \vert \sigma_i^z = +1 \rangle$ 
and we introduce the group $G$ generated by products of $A_s$ operators. 
Notice that one has to define elements $g \in G$ modulo the identity 
$\prod_s A_s = \Id$ in order for the inverse of $g$ to be uniquely defined 
(in which case, $g^{-1} = g$). The dimension (i.e., the number of elements) 
of $G$ is therefore $\vert G \vert = 2^{N^{(s)}-1}$, where $N^{(s)}$ is 
the number of sites on the lattice. 
A GS can then be written explicitly as: 
\beq
\vert \psi_0 \rangle 
= 
\frac{1}{\vert G \vert^{1/2}} 
  \sum_{g \in G} g \vert 0 \rangle 
. 
\label{eq: TC GS}
\eeq
%
%
%
%

\section{Choice of partitions}  
Here we are interested in computing the negativity of the system, in order to 
understand its relation to the topological correlations between two 
subsystems $\Ai$ and $\Aii$ without knowledge of the rest of the 
system $\B$ ($\S = \Ai \cup \Aii \cup \B$). 
Since $\N$ is a measure of separability of subsystems $\Ai$ and $\Aii$, 
one expects $\N=0$ if $\Ai$ and $\Aii$ are topologically trivial. 
This expectation is confirmed by the calculation below. 

\begin{figure}[!ht]
\includegraphics[width=0.8\columnwidth]{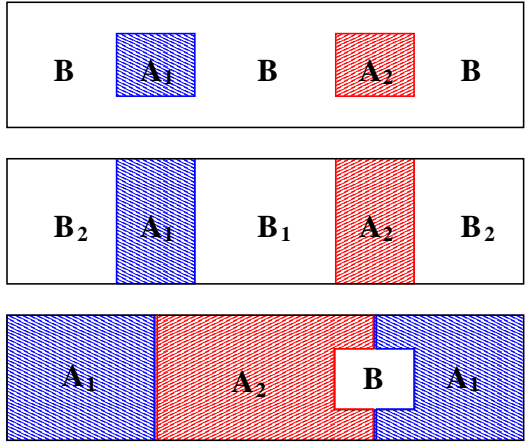}
\caption{
\label{fig: partitions} 
(Colour online) -- 
Examples of tri-partitions of the system into $\S = \Ai \cup \Aii \cup \B$. 
The top panel illustrates a topologically trivial choice for $\Ai$ and $\Aii$, 
which corresponds to vanishing negativity. Although subsystems 
$\Ai$ and $\Aii$ are no longer trivial in the middle panel, as they wind 
around the system in the vertical direction, the negativity remains zero. 
The bottom panel illustrates a choice of partitions with finite logarithmic 
negativity between $\Ai$ and $\Aii$, where $\E$ exhibits both a 
boundary as well as a topological contribution. 
}
\end{figure}

In order to understand the behaviour of the negativity, we consider 
progressively more involved choices of partitions, 
as illustrated in Fig.~\ref{fig: partitions}. 
This work will form the basis to understand how to identify the topological 
contribution in $\N$. 
%
%

\section{Negativity of the toric code}  
The density matrix of subsystem $\A = \Ai \cup \Aii$ is given by 
$\rho_\A = {\rm Tr}_\B \rho$, where 
$\rho = \vert\psi_0\rangle\langle\psi_0\vert$ is prepared in the pure 
state~\eqref{eq: TC GS}: 
\beq
\rho_\A 
= 
\frac{1}{\vert G \vert} 
\sum_{g,g' \in G} 
g_\A \vert 0_\A \rangle\langle 0_\A \vert g'_\A 
\; 
\langle 0_\B \vert g'_\B g_\B \vert 0_\B \rangle 
, 
\eeq
where we introduced the notation 
$\vert 0 \rangle = \vert 0_\A \rangle \otimes \vert 0_\B \rangle$ and 
$g = g_\A \otimes g_\B$. 
It is convenient to redefine $g' \to g g'$. Note that, for any given $g$, 
this mapping for $g'$ is 1-to-1 in $G$. The trace over $\B$ imposes then 
that $g'$ acts trivially on $\B$ 
($\langle 0_\B \vert g'_\B \vert 0_\B \rangle = 1$), i.e., 
$g' \in G_\A \equiv \{ g' \in G \vert g'_\B = \Id_\B \}$. 
We thus arrive at the expression: 
\beq
\rho_\A 
= 
\frac{1}{\vert G \vert} 
\sum_{g \in G} 
\sum_{g' \in G_\A} 
g_\A \vert 0_\A \rangle\langle 0_\A \vert g_\A g'_\A 
. 
\eeq

Next, we take the partial transpose over $\Aii$. Given that we can choose 
all elements of $\rho_\A$ to be real, the transpose is equivalent to the 
adjoint of the part of $\rho_\A$ acting on $\Aii$: 
\bea
\rho_\A^{T_2} 
&=& 
\frac{1}{\vert G \vert} 
\sum_{g \in G} 
\sum_{g' \in G_\A} 
\left(\vphantom{\sum}
g_{\Ai} \vert 0_{\Ai} \rangle\langle 0_{\Ai} \vert g_{\Ai} g'_{\Ai} 
\right)
\nonumber \\ 
&&\qquad\qquad\quad
\otimes
\left(\vphantom{\sum}
g_{\Aii} g'_{\Aii} \vert 0_{\Aii} \rangle\langle 0_{\Aii} \vert g_{\Aii} 
\right)
. 
\eea
Following Ref.~\onlinecite{Calabrese2012}, we want to obtain the trace norm 
of $\rho_\A^{T_2}$ with a replica approach as the analytic continuation 
for $n \to 1/2$ of ${\rm Tr} \left( \rho_\A^{T_2} \right)^{2n}$. Therefore, 
we need to compute 
\begin{widetext}
\bea
{\rm Tr} \left( \rho_\A^{T_2} \right)^{n} 
&=& 
\frac{1}{\vert G \vert^n} 
\sum_{g_1 \ldots g_n \in G} 
\sum_{g'_1 \ldots g'_n \in G_\A} 
\nonumber \\ 
&&
\langle 0_{\Ai} \vert g_{1\Ai} g_{2\Ai} g'_{1\Ai} \vert 0_{\Ai} \rangle
\ldots 
\langle 0_{\Ai} \vert g_{(n-1)\Ai} g_{n\Ai} g'_{(n-1)\Ai} \vert 0_{\Ai} \rangle
\langle 0_{\Ai} \vert g_{n\Ai} g_{1\Ai} g'_{n\Ai} \vert 0_{\Ai} \rangle
\nonumber \\ 
&&
\langle 0_{\Aii} \vert g_{1\Aii} g_{2\Aii} g'_{2\Aii} \vert 0_{\Aii} \rangle
\ldots
\langle 0_{\Aii} \vert g_{(n-1)\Aii} g_{n\Aii} g'_{n\Aii} \vert 0_{\Aii} \rangle
\langle 0_{\Aii} \vert g_{n\Aii} g_{1\Aii} g'_{1\Aii} \vert 0_{\Aii} \rangle
. 
\label{eq: Tr rhoAT2 n}
\eea
\end{widetext}

In general, the subgroup $G_\A \subset G$ decomposes into the product 
$G_{\Ai} \cdot G_{\Aii} \cdot G_{\Ai\Aii}$, where the quotient group 
$G_{\Ai\Aii} \equiv G_{\A}/(G_{\Ai}G_{\Aii})$ is defined as the set of 
$g \in G_\A$ that are equivalent up to the action of elements of $G_{\Ai}$ 
and $G_{\Aii}$. 
For disjoint, topologically trivial partitions (top panel in 
Fig.~\ref{fig: partitions}), $G_{\Ai\Aii} = \{\Id\}$ and 
$G_\A = G_{\Ai} \cdot G_{\Aii}$. 

This is no longer the case, for instance, with the choice of partitions in 
the middle panel in Fig.~\ref{fig: partitions}. 
Here $\A$ divides $\B$ into two disconnected portions $\B = \Bi \cup \Bii$. 
The product of all star operators $A_s$ acting on at least one spin in, 
say, $\Bi$ is an element $k \in G_\A$ (all spins in $\B$ are flipped either 
zero or two times). However, this element is special in that it 
\emph{cannot} be decomposed as product of elements of $G_{\Ai}$ and $G_{\Aii}$. 
[Equivalently if we had chosen the product $k'$ of all stars acting on at 
least one spin of $\Bii$; however, the product $k k'$ \emph{is} an element 
of $G_{\Ai} \cdot G_{\Aii}$, and therefore $k$ and $k'$ are identified in 
the quotient group $G_{\A}/(G_{\Ai}G_{\Aii})$.] 
One can verify that this is the only operation in $G_\A$ that is not a 
product of elements in $G_{\Ai}$ and $G_{\Aii}$. 
Therefore, $G_{\Ai\Aii} = \{ \Id, k \}$. 
In the case where $\B$ has more than 2 disconnected components, another 
independent element of $G_{\Ai\Aii}$ can be found per additional 
component (cf. Refs.~\onlinecite{Hamma2005},\onlinecite{Castelnovo2008}). 

If $\Ai$ and $\Aii$ share a common edge, as is the case in the bottom 
panel of Fig.~\ref{fig: partitions}, then the system allows for single 
star operators acting simultaneously (and only) on the two subsystems. 
In this case, each boundary star operator is an additional generator of 
$G_{\Ai\Aii}$. 

In general, for each element $g'_\ell \in G_\A$ there exists a unique 
decomposition $g'_\ell = \og_\ell \oog_\ell \theta_\ell$, with 
$\og_\ell \in G_{\Ai}$, 
$\oog_\ell \in G_{\Aii}$, and $\theta_\ell \in G_{\Ai\Aii}$. 
Accordingly, one can write the first $n-1$ expectation values in 
each row of Eq.~\eqref{eq: Tr rhoAT2 n} as: 
\bea
&&
\langle 0_{\Ai} \vert 
  g_{(\ell-1)\Ai} g_{\ell\Ai} \og_{(\ell-1)\Ai} \theta_{\ell-1} 
\vert 0_{\Ai} \rangle
\\
&&
\langle 0_{\Aii} \vert 
  g_{(\ell-1)\Aii} g_{\ell\Aii} \oog_{\ell\Aii} \theta_{\ell} 
\vert 0_{\Aii} \rangle
, 
\eea
for $\ell=2,\ldots,n$. It is then convenient to redefine 
$
g_\ell \to \tg_{\ell} 
\equiv 
g_{\ell-1} \, g_\ell \, \og_{\ell-1} \, \oog_{\ell}
$, 
$\ell=2,\ldots,n$. 
Once again, this is a 1-to-1 mapping $g_\ell \to \tg_{\ell}$ in $G$ given 
$g_{\ell-1}$, $\og_{\ell-1}$, and $\oog_{\ell}$. 
Upon fixing the first term, say $\tg_1 = g_1$, the new definition is nothing 
but a re-labelling of $n-1$ mute indices in the summation over 
$g_1 \ldots g_n$. This simplifies the above expectation values to: 
\bea
&&
\langle 0_{\Ai} \vert 
  \tg_{\ell\Ai} \theta_{\ell-1} 
\vert 0_{\Ai} \rangle
\nonumber\\
&&
\langle 0_{\Aii} \vert 
  \tg_{\ell\Aii} \theta_{\ell} 
\vert 0_{\Aii} \rangle
, 
\label{eq: first n-1 terms}
\eea
which vanish \emph{unless} $\tg_{\ell} \theta_{\ell-1}$ acts trivially on 
$\Ai$ \emph{and} $\tg_{\ell} \theta_{\ell}$ acts trivially on $\Aii$. 

The last expectation value in each of the two rows in 
Eq.~\eqref{eq: Tr rhoAT2 n} needs to be dealt with separately. 
Upon combining the chain of mappings $g_\ell \to \tg_{\ell}$, one obtains 
$
\tg_n 
= 
\left(
\prod_{\ell = 1}^{n-1} \tg_\ell \, \og_\ell
\right)
\left(
\prod_{\ell = 2}^{n} \oog_\ell
\right)
g_n 
$
and the remaining two expectation values can be written as 
\bea
&&
\langle 0_{\Ai} \vert 
  \left(
    \prod_{\ell = 1}^{n-1} \tg_{\ell\Ai} \, \og_{\ell\Ai}
  \right)
  \tg_{n\Ai} \tg_{1\Ai} \og_{n\Ai} \theta_n
\vert 0_{\Ai} \rangle
\nonumber\\ 
&&
\langle 0_{\Aii} \vert 
  \left(
    \prod_{\ell = 1}^{n-1} \tg_{\ell\Aii} 
  \right)
  \left(
    \prod_{\ell = 2}^{n} \oog_{\ell\Aii}
  \right)
	\tg_{n\Aii} \tg_{1\Aii} \oog_{1\Aii} \theta_1
\vert 0_{\Aii} \rangle
, 
\nonumber
\eea
where we used the fact that $\oog_{\ell\Ai} = \Id_{\Ai}$ and 
$\og_{\ell\Aii} = \Id_{\Aii}$, by definition. 
After straightforward algebraic manipulation, we obtain 
\bea
&&
\langle 0_{\Ai} \vert 
  \left(
    \prod_{\ell = 2}^{n} \tg_{\ell\Ai}
  \right)
  \left(
    \prod_{\ell = 1}^{n} \og_{\ell\Ai}
  \right)
	\theta_n
\vert 0_{\Ai} \rangle
\nonumber\\ 
&&
\langle 0_{\Aii} \vert 
  \left(
    \prod_{\ell = 2}^{n} \tg_{\ell\Aii}
  \right)
  \left(
    \prod_{\ell = 1}^{n} \oog_{\ell\Aii}
  \right) 
	\theta_1
\vert 0_{\Aii} \rangle
. 
\nonumber
\eea

Notice that the dependence on $\tg_1$ has disappeared. 
We can further simplify these expressions using 
Eqs.~\eqref{eq: first n-1 terms}. Since each $\tg_{\ell} \theta_{\ell-1}$ 
acts trivially on $\Ai$ and $\tg_{\ell} \theta_{\ell}$ acts trivially on 
$\Aii$ for $\ell = 2,\ldots,n$, then this is true of their product. 
Therefore, 
$\prod_{\ell = 2}^n \tg_{\ell\Ai} = \prod^{n-1}_{\ell = 1} \theta_\ell$ 
and 
$\prod_{\ell = 2}^n \tg_{\ell\Aii} = \prod^{n}_{\ell = 2} \theta_\ell$. 
Substituting into the above equations, they reduce to 
\bea
&&
\langle 0_{\Ai} \vert 
  \left(
    \prod_{\ell = 1}^{n} \theta_{\ell}
  \right)
  \left(
    \prod_{\ell = 1}^{n} \og_{\ell\Ai}
  \right)
\vert 0_{\Ai} \rangle
\nonumber\\ 
&&
\langle 0_{\Aii} \vert 
  \left(
    \prod_{\ell = 1}^{n} \theta_{\ell}
  \right)
  \left(
    \prod_{\ell = 1}^{n} \oog_{\ell\Aii}
  \right) 
\vert 0_{\Aii} \rangle
. 
\nonumber
\eea
%
%
%
%
%
%
By definition, products of $\theta_\ell \in G_{\Ai\Aii}$ other than the 
identity cannot be decomposed into products of $\og_\ell \in G_{\Ai}$ and 
$\oog_\ell \in G_{\Aii}$. Therefore, the above equations separately imply 
that $\prod_{\ell = 1}^{n} \theta_{\ell} = \Id$, 
$\prod_{\ell = 1}^{n} \og_{\ell\Ai} = \Id$, and 
$\prod_{\ell = 1}^{n} \oog_{\ell\Aii} = \Id$. 
We thus arrive at the expression: 
\begin{widetext}
\bea
{\rm Tr} \left( \rho_\A^{T_2} \right)^{n} 
&=& 
\frac{\vert G_{\Ai} \vert^{n-1} \vert G_{\Aii} \vert^{n-1}}
     {\vert G \vert^{n-1}} 
\sum_{\theta_1 \dots \theta_n \in G_{\Ai\Aii}} 
\prod_{\ell = 2}^{n}
\left(
\sum_{\tg_\ell \in G} 
\langle 0_{\Ai} \vert 
  \tg_{\ell\Ai} \theta_{\ell-1} 
\vert 0_{\Ai} \rangle
\langle 0_{\Aii} \vert 
  \tg_{\ell\Aii} \theta_{\ell} 
\vert 0_{\Aii} \rangle
\right)
\langle 0 \vert 
  \prod_{\ell = 1}^n \theta_{\ell} 
\vert 0 \rangle
. 
\nonumber \\ 
\label{eq: Tr rhoAT2 n bis}
\eea
\end{widetext}
%
%

The term in round brackets acts as a projector: it vanishes identically 
unless the action of $\theta_{(\ell-1)\Ai}\,\theta_{\ell\Aii}$ can be matched 
by $\tg_{\ell\A}$. In that case, the expectation values equal $1$, 
$\tg_{\ell}$ is uniquely selected over subsystem $\A$, and the summation 
over $\tg_\ell \in G$ contributes a factor $\vert G_\B \vert$. 

Let us consider for example the three cases illustrated in 
Fig.~\ref{fig: partitions}. 
The case in the top panel corresponds to $G_{\Ai\Aii} = \{\Id\}$; the 
summation over $\theta_\ell$ is not present and the result greatly simplifies. 
As discussed in more detail in the Appendix, this choice of 
partition leads to vanishing negativity. 

In the middle panel of Fig.~\ref{fig: partitions}, the partitions $\Ai$ and 
$\Aii$ are no longer topologically trivial (in the sense that they wind 
around the system in one direction). Here $G_{\Ai\Aii} = \{\Id,k\}$, 
where $k$ is given by the product of all star operators acting on $\Bi$. 
Nonetheless, after some consideration one can see that 
the expectation values between round brackets in 
Eq.~\eqref{eq: Tr rhoAT2 n bis} do not impose any limitations on the choice 
of $\theta_{\ell}$. 
The only difference to the previous case is that the summation over 
$\theta_\ell$ subject to the constraint 
$
\langle 0 \vert 
  \prod_{\ell = 1}^n \theta_{\ell} 
\vert 0 \rangle
$ 
results in an additional factor 
$
2^{(\vert G_{\Ai\Aii} \vert - 1)(n-1)} 
= 
2^{n-1}
$, 
and once again the negativity vanishes. 

Let us finally consider the partition in the bottom panel 
in Fig.~\ref{fig: partitions}. 
Subsystem $\B$ has only one component; however, subsystems $\Ai$ and $\Aii$ 
have now two direct boundaries with one another (recall that the figure has 
periodic boundary conditions). 
The group $G_{\Ai\Aii}$ is generated by all the star operators acting 
simultaneously (and exclusively) on $\Ai$ and $\Aii$. One can also define 
the operator $k$ given by the product of all the stars acting on at least 
one spin in $\B$. 
However, the product of all $\Ai-\Aii$ boundary star operators times the 
operator $k$ is nothing but the product of all stars acting solely on $\Ai$ 
and $\Aii$, which is an operator that belongs to $G_{\Ai} \cdot G_{\Aii}$. 
Therefore, $k$ is in fact equivalent to the product of all 
$\Ai-\Aii$ boundary stars, and 
$\vert G_{\Ai\Aii} \vert = 2^{N^{(s)}_{\Ai-\Aii}}$. 
In the following, it is convenient to think of the group $G_{\Ai\Aii}$ as 
the group generated by $k$ and by products of $\Ai-\Aii$ boundary star 
operators defined modulo the product of \emph{all} $\Ai-\Aii$ boundary star 
operators. 

One can verify that the action of $k$ on the expectation values in 
Eq.~\eqref{eq: Tr rhoAT2 n bis} is immaterial, much as is the case for the 
partition in the middle panel of Fig.~\ref{fig: partitions} considered earlier. 
On the contrary, products of $\Ai-\Aii$ boundary star operators 
(identified modulo the product of all of them) play a crucial role. 
If $\theta_{\ell-1}$ and $\theta_{\ell}$ differ in this respect, it is then 
not possible to find any $\tg_\ell \in G$ such that 
$\tg_{\ell\A} = \theta_{(\ell-1)\Ai}\,\theta_{\ell\Aii}$. 
The expectation values in Eq.~\eqref{eq: Tr rhoAT2 n bis} vanish unless 
all $\theta_\ell$ have the same contribution of products of $\Ai-\Aii$ 
boundary star operators. This leads to a significant 
difference in the behaviour of ${\rm Tr} \left( \rho_\A^{T_2} \right)^{n}$ 
for even or odd $n$. Indeed, if $n$ is odd, the product of $\theta_\ell$ 
can be the identity only if all $\theta_\ell = \Id$. Vice versa, if $n$ is 
even, the product always equals the identity irrespective of the choice of 
$\theta_\ell$. As a result, Eq.~\eqref{eq: Tr rhoAT2 n bis} becomes 
%
%
\bea
{\rm Tr} \left( \rho_\A^{T_2} \right)^{n} 
&=& 
f(n)
\left[
  \frac{2 \vert G_{\Ai} \vert \vert G_{\Aii} \vert \vert G_\B \vert}
       {\vert G \vert} 
\right]^{n-1} 
\label{eq: Tr rhoAT2 n final}
\eea
%
%
where $f(n) = 1$ for $n$ odd, and $f(n) = 2^{N^{(s)}_{\Ai-\Aii} - 1}$ if 
$n$ is even. 

The behaviour of the factor $f(n)$ leads to a different analytic 
continuation of ${\rm Tr} \left( \rho_\A^{T_2} \right)^{n}$ for even and 
odd $n$. 
If we follow the odd sequence, then 
${\rm Tr} \left( \rho_\A^{T_2} \right)^{n} \to 1$ for $n \to 1$, 
as expected for a quantity that converges to the sum of the eigenvalues of 
a density matrix operator. 
If instead we follow the even sequence, we obtain 
$
\| \rho_\A^{T_2} \|_1 
= 
\lim_{n \to 1/2} {\rm Tr} \left( \rho_\A^{T_2} \right)^{2n} 
= 
2^{N^{(s)}_{\Ai-\Aii} - 1}
$ 
and therefore $\N = (2^{N^{(s)}_{\Ai-\Aii} - 1} - 1)/2$, 
$\E = (N^{(s)}_{\Ai-\Aii} - 1 )\ln 2$. 

The leading behaviour is akin to the well-known area law observed in the 
scaling of the entanglement entropy. 
The correction of order one is instead universal and it is directly related 
to the topological entropy $\gamma$ of the quantum 
system~\cite{Levin2006,Kitaev2006}. 
%
%

\section{Conclusions}  
We performed an exact calculation of the negativity for the toric code 
model using different choices of partitions. 
We find that the negativity has a leading area-law contribution, if the 
subsystems are in direct contact with one another, as expected in a 
zero-range correlated model. We also find a topological contribution 
reflecting the topological nature of the quantum state, provided that 
subsystem $\B$ (which is traced out) does not span the system 
in either direction. This topological contribution is 
directly related to the topological entropy $\gamma$. 
%
%
As in the case of the von Neumann entropy, a direct measure of 
$\gamma$ likely requires either a subtraction scheme or finite-size 
extrapolation~\cite{Levin2006,Kitaev2006}. 

It is interesting to recall that other approaches to probe the topological 
nature of the system, typically based on the von Neumann entropy, 
yield a non-vanishing value of $\gamma$ also for classical topologically 
ordered systems~\cite{Castelnovo2007} (e.g., in the 8-vertex model). 
A straightforward calculation (see Appendix) shows that 
the negativity vanishes identically in the classical 8-vertex model, 
consistently with the expectation that $\N > 0$ is a measure of quantum 
entanglement only. 

Comparing the calculations in this paper with the work in 
Refs.~\onlinecite{Castelnovo2007finT,Castelnovo2008finT}, one expects 
that the topological contribution to the negativity vanishes 
in the 2D toric code at any finite temperature in the thermodynamic 
limit. 
Contrary to the behaviour of the topological entropy, this ought 
to be true also for the toric code in 3D, which reduces to a classical 
$\mathbb{Z}_2$ gauge theory at finite temperature. 
It will be interesting to see whether the finite size behaviour of the 
negativity at finite temperature is able to discern the low temperature 
phase of the classical $\mathbb{Z}_2$ gauge theory from the trivial 
paramagnetic phase at high temperature, despite the fact that $\N$ 
vanishes in both cases in the thermodynamic limit. 
Finally, only in the 4D toric code one might expect quantum topological 
correlations to actually survive at finite temperature, and thus $\N > 0$ 
for $T > 0$. 

To some extent the toric code is a rather special example of topological 
order with precisely `zero-ranged' local correlations. It will be interesting 
to see extensions of the calculation of the negativity to other topologically 
non-trivial states in dimensions larger than one. 
One could perhaps start from perturbations of the toric code introduced via 
stochastic matrix form decomposition~\cite{Castelnovo2008}, where the GS 
wavefunction is known exactly throughout the phase diagram. 
These perturbations introduce finite correlations and eventually drive the 
system across a so-called conformal critical point. 
It may also be possible to study the behaviour of the negativity at such 
critical points by means of conformal field theoretic 
techniques~\cite{Fradkin2006,Castelnovo2007,Oshikawa2010}. 
This work could lead the way to the much more challenging and interesting 
question of investigating the behaviour of the negativity in quantum Hall 
states and other topologically ordered phases of matter. 

After this work was completed, private communication with G.~Vidal revealed 
that, together with A.~Lee, they had independently arrived at similar 
results~\cite{Lee2013}. 
The author is deeply indebted to G.~Vidal for spotting an inconsistency in 
the first version of this manuscript. 
%
%

\section*{Acknowledgments} 
This paper owes its existence to P.~Calabrese, who introduced the author to 
the concept of negativity and stimulated interest in an exact calculation 
for the toric code model. 
This work was supported by EPSRC Grant EP/K028960/1. 
%
%

\section{Appendix} 

\subsection{Classical 8-vertex model}

The classical 8-vertex model is a combinatorial problem of arrows on the 
bonds of the square lattice, with the hard constraint that the number of 
incoming arrows at every vertex is even (counting $0$ as an even number). 
Taking advantage of the bipartite nature of the lattice, we can define 
arrows going from sublattice A to sublattice B as positive spins, and all 
others are negative. This establishes a 1-to-1 mapping between 8-vertex 
configurations and $\sigma^z$ tensor product states that minimise the energy 
of the plaquette term in the toric code 
Hamiltonian~\eqref{eq: TC Hamiltonian}. 
All 8-vertex configurations can be obtained from a reference configuration, 
say the spin polarized $\vert 0 \rangle$, by acting with elements of $G$. 

The partition function of the 8-vertex model can thus be written in ket-bra 
notation as 
\beq
\rho 
= 
\frac{1}{\vert G \vert} 
\sum_{g \in G} 
g \vert 0 \rangle\langle 0 \vert g 
. 
\eeq
Taking the trace over $\B$ is straightforward, since $g_\B^2 = \Id$, 
and $\rho_\A$ remains diagonal: 
\beq
\rho_\A 
= 
\frac{1}{\vert G \vert} 
\sum_{g \in G} 
g_\A \vert 0_\A \rangle\langle 0_\A \vert g_\A 
= \rho_\A^{T_2}
. 
\eeq

In order to use the replica approach, we need to compute 
%
%
\bea
{\rm Tr} \left( \rho_\A^{T_2} \right)^{n} 
&=& 
\frac{1}{\vert G \vert^n} 
\sum_{g_1 \ldots g_n \in G} 
\langle 0_{\A} \vert g_{1\A} g_{2\A} \vert 0_{\A} \rangle
\ldots 
\nonumber \\
&&
\langle 0_{\A} \vert g_{(n-1)\A} g_{n\A} \vert 0_{\A} \rangle
\langle 0_{\A} \vert g_{n\A} g_{1\A} \vert 0_{\A} \rangle
. 
\nonumber
\eea
%
%
It is convenient to redefine 
$g_\ell \to \og_\ell \equiv g_{\ell - 1} \,g_\ell$, 
$\ell = 2,\ldots,n$, with the choice $\og_1 \equiv g_1$. 
All expectation values simplify to 
$\langle 0_{\A} \vert \og_{\ell\A} \vert 0_{\A} \rangle$ 
for $\ell = 2,\ldots,n$ except for the last one, where 
the chain of mappings leads to 
$\og_n \equiv \prod_{\ell = 1}^{n-1} \og_\ell \, g_n$ 
and therefore to the expectation value 
$
\langle 0_{\A} \vert 
\prod_{\ell = 2}^{n} \og_{\ell\A} 
\vert 0_{\A} \rangle
$. 
Once again, the dependence on $\og_1$ has disappeared, and the expectation 
values impose that all other $\og_\ell$ acts trivially on $\A$ 
($\og_\ell \in G_\B$ for $\ell = 2,\ldots,n$). 
As a results, we obtain: 
%
%
\bea
{\rm Tr} \left( \rho_\A^{T_2} \right)^{n} 
&=& 
\frac{\vert G_\B \vert^{n-1}}{\vert G \vert^{n-1}} 
, 
\nonumber\\
&=&
\left[
  2^{-N^{(s)}_\A - N^{(s)}_{\delta\A} + 2} 
\right]^{n-1}
, 
\label{eq: Tr rhoAT2 n classical}
\eea
%
%
where $\vert G \vert = 2^{N^{(s)}-1}$ and 
$\vert G_\B \vert = 2^{N^{(s)}_\B+1}$ (the latter is due to the fact that 
the product of all stars acting on at least one spin in $\Ai$ is an element 
of $G_\B$ that cannot be written as a product of stars belonging to $G_\B$, 
see e.g., Ref.~\onlinecite{Castelnovo2008}). 
Here $N^{(s)} = N^{(s)}_\A + N^{(s)}_\B + N^{(s)}_{\delta\A}$, where 
$N^{(s)}_{\delta\A}$ is the number of star operators acting simultaneously on 
spins in $\A$ and $\B$. 

The three contributions in the final expression of 
Eq.~\eqref{eq: Tr rhoAT2 n classical} correspond, respectively, to the 
classical entropy (scaling with the volume of subsystem $\A$), the area 
law, and a classical topological contribution. 

Due to the diagonal nature of the density matrix, the topologically trivial 
vs non-trivial character of $\Ai$ and $\Aii$ does not play a role 
(whereas the fact that $\B$ is non-trivial plays a crucial part in this case). 
As a result, the even and odd analytic continuations of 
${\rm Tr} \left( \rho_\A^{T_2} \right)^{n}$ coincide and in the limit 
$n \to 1$ we obtain $\| \rho_\A^{T_2} \|_1 = 1$, and therefore $\N = 0$, 
$\E = 0$. 


\subsection{Topologically trivial partitions}

It is interesting to briefly consider what happens to the calculation of 
${\rm Tr} \left( \rho_\A^{T_2} \right)^{n}$ when the partitions $\Ai$ 
and $\Aii$ are disjoint and topologically trivial. 
As discussed earlier, $\theta_\ell = \Id$ is the only choice throughout 
Eq.~\eqref{eq: Tr rhoAT2 n bis}. 

In this case, $G_\A = G_{\Ai} \cdot G_{\Aii}$ and 
$G_\B \equiv G_{\A^c} = G_{\Ai^c} \cap G_{\Aii^c}$. Therefore, 
\bea
{\rm Tr} \left( \rho_\A^{T_2} \right)^{n} 
&=& 
\left[ 
\frac{
     \vert G_\B \vert
		 \vert G_{\Ai} \vert
		 \vert G_{\Aii} \vert
     }{\vert G \vert} 
\right]^{n-1}
\nonumber \\ 
&=& 
\left[ 
2^{-N^{(s)}_{\delta\A} + 2}
\right]^{n-1}
, 
\label{eq: Tr rhoAT2 n trivial}
\eea
where we used the fact that 
$N^{(s)} - N^{(s)}_{\Ai} - N^{(s)}_{\Aii} - N^{(s)}_\B = N^{(s)}_{\delta\A}$. 
Once again we recognise the leading area law and a topological contribution. 
However, the latter arises from the topologically non-trivial nature of $\B$ 
rather than $\Ai$ or $\Aii$, and it does not contribute to the negativity 
between the latter two subsystems. 
Indeed, the even and odd analytic continuations of 
${\rm Tr} \left( \rho_\A^{T_2} \right)^{n}$ coincide, and in the limit 
$n \to 1$ we obtain $\| \rho_\A^{T_2} \|_1 = 1$, and therefore $\N = 0$, 
$\E = 0$.
%
%


\end{document}